\documentclass[11pt]{article}
\usepackage{amsmath,amssymb,color,graphics,epsfig,cite}

\usepackage{amsfonts}

\newcommand{\be}{\begin{equation}}
\newcommand{\ee}{\end{equation}}
\newcommand{\bea}{\setlength\arraycolsep{2pt} \begin{eqnarray}}
\newcommand{\eea}{\end{eqnarray}}
\newcommand{\nn}{\nonumber}

\def\ft#1#2{{\textstyle{\frac{\scriptstyle #1}{\scriptstyle #2} } }}
\def\fft#1#2{{\frac{#1}{#2}}}

\def\0{{\sst{(0)}}}
\def\1{{\sst{(1)}}}
\def\2{{\sst{(2)}}}
\def\3{{\sst{(3)}}}
\def\4{{\sst{(4)}}}
\def\5{{\sst{(5)}}}
\def\6{{\sst{(6)}}}
\def\7{{\sst{(7)}}}
\def\8{{\sst{(8)}}}
\def\sst#1{{\scriptscriptstyle #1}}

\begin{document}

\begin{center}
{\Large {\bf Quadratic Curvature Correction to 5D Myers-Perry Metric}}

\vspace{20pt}

Liang Ma and H. L\"u
		
\vspace{10pt}

{\it Center for Joint Quantum Studies, Department of Physics,\\
School of Science, Tianjin University, Tianjin 300350, China }

\vspace{40pt}

\underline{ABSTRACT}
\end{center}

We consider quadratic curvature perturbation to the Myers-Perry black hole in five dimensions at the linear level in the coupling constant. The solution can then be solved order by order in terms of two dimensionless angular momentum parameters up to an arbitrary order. We present the results up to tenth order. The perturbed solution allows us to obtain the higher-derivative correction to the black hole thermodynamics, which we find is in complete agreement with the Reall-Santos method.

\vfill{maliang0@tju.edu.cn \ \ \  mrhonglu@gmail.com}

\thispagestyle{empty}
\pagebreak



\section{Introduction}

Since Einstein formulated General Relativity (GR) over a century ago, it has withstood increasingly remarkable tests, providing an accurate and reliable description of gravitational interactions. However, GR is far from the ultimate theory and has long faced fundamental incompatibility with quantum mechanics. With the development of quantum gravity frameworks such as string theory and the formulation of effective field theory (EFT), it becomes necessary to introduce higher-derivative corrections to GR \cite{Stelle:1976gc,Stelle:1977ry}, and even to explore gravitational interactions in higher spacetime dimensions. A notable example is the four-derivative correction to the heterotic supergravity \cite{Bergshoeff:1989de}.  In this paper, we focus on pure gravity and the action of a general EFT takes the form
\bea
S=\frac{1}{16\pi}\int d^Dx\sqrt{-g}\big(\mathcal{L}_{\mathrm{EH}}+\Lambda_c^{-2}\mathcal{L}_{4\partial}
+\Lambda_c^{-4}\mathcal{L}_{6\partial}+\cdots\big)\label{D-dim EFT}
\eea
The leading $\mathcal{L}_{\mathrm{EH}}=R$ corresponds to the Einstein-Hilbert term, whose variation yields the standard two-derivative Einstein field equations. The higher-derivative contributions $\mathcal{L}_{4\partial}$, $\mathcal{L}_{6\partial}$ etc., arise from integrating out heavy degrees of freedom. The ultraviolet cutoff scale $\Lambda_c$ ensures that the effective theory \eqref{D-dim EFT} remains unitary below this scale and hence delineates the range of validity of the EFT.

Black hole physics plays a central role in the study of GR. Owing to their enormous masses and extremely strong gravitational fields, the physical phenomena associated with black holes lie far beyond the scope of Newtonian gravity, and can be consistently described only within the GR framework. In this sense, black holes constitute one of the most profound and characteristic predictions of GR. The development of black hole thermodynamics \cite{Bekenstein:1973ur,Hawking:1975vcx} represents an crucial step toward bridging classical gravity and quantum physics. Within the framework of semiclassical quantum field theory in curved spacetime \cite{Hawking:1975vcx,Gibbons:1976ue}, black holes are found to behave like conventional thermodynamic system: they emit thermal radiation, and are characterized by the well-defined quantities such as temperature and entropy. Hence, investigating the effects of higher-derivative connections on black hole thermodynamics not only refines and extends the thermodynamic description of black holes, but also provides a valuable probe into how the ultraviolet modifications to gravity manifest themselves within the classical theory of GR.

Exact black hole solutions provide a fundamental basis for investigating black hole thermodynamics and related physical processes.\footnote{Recently it was shown that, for string or string-inspired theories, the complete set of thermodynamic quantities can be obtained directly from the theory, without solving for field equations \cite{Lu:2025eub,Yang:2025rud}.} However, due to the highly nonlinear nature of the Einstein field equations, such solutions are exceedingly rare. Compared to the spherically symmetric, static Schwarzschild and Reissner-Nordstr\"om (RN) black holes, the axisymmetric rotating Kerr solution is considerably more involved. This complexity becomes even more pronounced in higher dimensions. While higher-dimensional generalizations of the Schwarzschild or the RN black holes are relatively straightforward, rotating black holes grow substantially more complicated as the spacetime dimensions increase. This is because the number of independent (orthogonal) rotation planes rises with dimensions, leading to multiple independent angular momenta as well as increasing cohomogeneity, as demonstrated by the Myers-Perry (MP) solutions \cite{Myers:1986un} and their generalizations to include a cosmological constant \cite{Hawking:1998kw,Gibbons:2004uw,Gibbons:2004js}.

Due to the intrinsic complexity of rotating black holes even in two-derivative GR, higher-derivative corrections to rotating solutions are rarely accessible in closed analytic form, except for a few exceptional cases \cite{Ma:2020xwi}. As a result, such corrections are typically investigated either numerically or within the slow-rotation approximation \cite{Cardoso:2018ptl}. Recently, a new approach based on slow-rotation expansion has been proposed in \cite{Cano:2019ore}, in which the solution can be constructed perturbatively to arbitrary higher order in the dimensionless rotation parameter $\chi = a/\mu$. This enables a highly accurate description of the black holes even when the angular momentum is not parametrically small. When implementing this method, 
the expansion must be performed in a specific order: one should first expand in the cutoff scale $\Lambda_c$ and solve the two-derivative equations, and only subsequently perform the rotation $\chi$ expansion for the higher-derivative corrections. A similar approach is presented in \cite{Fernandes:2025vxg}, which follows the same scheme as \cite{Cano:2019ore}, but treats the higher-derivative sector using a pseudospectral collocation method. To date, this method has been widely applied to the study of black hole multipole moments \cite{Cano:2022wwo,Ma:2024ulp}, quasinormal modes \cite{Cano:2023jbk,Cano:2024wzo,Cano:2024jkd,Cano:2024ezp}, and black hole quasi-periodic oscillations \cite{Allahyari:2025bvf}, among other topics.

The success of the method fundamentally relies on the axisymmetry of the Kerr black hole, which is required to preserve by the higher-derivative corrections \eqref{D-dim EFT}. In other words, $\partial_t$ and $\partial_\varphi$ remain Killing vectors of the corrected metric. As a consequence, a perturbative ansatz can still be consistently constructed that depends only on the radial coordinate $r$ and the latitude coordinate $\theta$. In this sense, the five-dimensional MP black hole closely resembles its four-dimensional counterpart. The extension from four to five dimensions introduces an additional azimuthal angle, leading to two independent angular coordinates $(\varphi, \psi)$, associated with two orthogonal rotating planes. However, the metric remains cohomogeneity two, with a radial coordinate and a latitude coordinate. This property significantly simplifies the construction of the perturbative ansatz.

The paper is organized as follows. In section 2, we start from the five-dimensional MP solution and construct the perturbative solution under the quadratic curvature correction.

The analysis involves two independent perturbative expansions. One is an expansion in higher-derivative coupling constant $\alpha$, while the other is an expansion in the two dimensionless angular momenta characterizing the rotating black hole. The solution is obtained to linear order in higher-derivative coupling, but can be constructed to arbitrarily higher in angular momenta. In section 3, we derive the complete set of the perturbed thermodynamic quantities and verify the first law. We conclude the paper in section 4.

It is worth noting that if one is interested only in deriving the quadratic curvature correction to the MP black hole thermodynamics,the Reall-Santos (RS) method can be employed without explicitly constructing the perturbative solution \cite{Reall:2019sah}. The method has been further generalized to $n$'th order perturbation where only the $(n-1)$'th order solution is needed \cite{Ma:2023qqj}. However, the RS method fundamentally relies on the assumption that perturbative solutions exist. This work provides an explicit confirmation of this assumption.

\section{5D MP solution and the higher-derivative correction}

In this section, we first review the thermodynamic properties of the five-dimensional MP black hole. We then consider Einstein gravity extended with quadratic curvature invariant and construct the perturbative solution.

\subsection{A brief review of 5D Kerr black hole}

Five-dimensional rotating black hole is a particular member of the MP solutions in diverse dimensions. We follow the notation of \cite{Hawking:1998kw}, where the five-dimensional rotating AdS black hole was constructed. By setting the cosmological constant to zero, one recovers the MP metric:
\bea
ds^2&=&-\frac{\Delta_r}{\rho^2}\big(dt-a\sin^2\theta d\varphi-b\cos^2\theta d\psi\big)^2+
\frac{\sin^2\theta}{\rho^2}\big[adt-(r^2+a^2)d\varphi\big]^2\cr
&&+\frac{\cos^2\theta}{\rho^2}\big[bdt-(r^2+b^2)d\psi\big]^2+\frac{\rho^2}{\Delta_r}dr^2+\rho^2d\theta^2\cr
&&+\frac{1}{r^2\rho^2}\Big[
ab dt-b(r^2+a^2)\sin^2\theta d\varphi-a(r^2+b^2)\cos^2\theta d\psi
\Big]^2\,,\nn\\
\Delta_r&=&\frac{(r^2+a^2)(r^2+b^2)}{r^2}-2\mu\,,\qquad \rho^2=r^2+a^2\cos^2\theta+b^2\sin^2\theta\,.\label{5D Kerr}
\eea
The metric is parameterized by three independent integration constants $(\mu,a,b)$, describing the mass and two independent angular momenta, along the two azimuthal angles $(\varphi,\psi)$. The black hole horizon $r_h$ is located at the largest real root of $\Delta_r(r_h) = 0$. The complete set of the thermodynamic quantities is given by
\bea
M_0&=&\frac{3 \pi   }{4}\mu\,,\qquad T_0=\frac{r_h^4-a^2 b^2}{2 \pi  r_h (r_h^2+a^2) (r_h^2+b^2)}\,,\qquad S_0=\frac{\pi^2  (r_h^2+a^2) (r_h^2+b^2)}{2r_h}\,,\cr
\Omega_{a,0}&=&\frac{a}{r_h^2+a^2}\,,\qquad J_{a,0}=\frac{\pi}{2}\mu a\,,\qquad \Omega_{b,0}=\frac{b}{r_h^2+b^2}\,,\qquad J_{b,0}=\frac{\pi}{2}\mu b\,.
\label{black hole thermodynamic}
\eea
Here, we use the subscript 0 to denote the thermodynamic quantities at leading order, which satisfy the first law of black hole thermodynamics
\bea
\delta M_0=T_0\delta S_0+\Omega_{a,0}\delta J_{a,0}+\Omega_{b,0}\delta J_{b,0}\,.\label{first law leading}
\eea
The Gibbs free energy associated with the Euclidean action is
\bea
G_0=M_0-T_0S_0-\Omega_{a,0} J_{a,0}-\Omega_{b,0} J_{b,0}=\frac{\pi  (r_h^2+a^2) (r_h^2+b^2)}{8 r_h^2}\,.
\eea

\subsection{Higher-derivative corrections and perturbed metric ansatz}

In $D=5$, we focus on the next-to-leading order correction in \eqref{D-dim EFT}, corresponding to the quadratic term
\be
\mathcal{L}=\mathcal{L}_{\mathrm{EH}}+\alpha\mathcal{L}_{4\partial}\,,\qquad
\mathcal{L}_{\mathrm{EH}}=R\,,\qquad \mathcal{L}_{4\partial}=R_{\mu\nu\rho\sigma}R^{\mu\nu\rho\sigma}\,.\label{quadratic}
\ee
The most general quadratic gravity in five dimensions also includes terms proportional to $R_{\mu\nu}R^{\mu\nu}$ and $R^2$. However, since the MP black hole \eqref{5D Kerr} is Ricci flat, these two terms do not contribute to the equation of motion at the order considered here. The modified Einstein equation is given by
\be
\mathcal{E}_{\mu\nu}\equiv
P_{(\mu}{}^{\alpha\beta\gamma}R_{\nu)\alpha\beta\gamma}-\ft{1}{2}g_{\mu\nu}\mathcal{L}
+2\nabla^{\alpha}\nabla^{\beta}P_{(\mu|\alpha|\nu)\beta}
=0\,,\label{eom}
\ee
where the four-index tensor $P_{\mu\nu\rho\sigma}$ is
\bea
P_{\mu\nu\rho\sigma} &\equiv &  \fft{\partial {\cal L}}{\partial R^{\mu\nu\rho\sigma}} =P_{E,\mu\nu\rho\sigma} + P_{4\partial,\mu\nu\rho\sigma}\,,\nn\\
P_{E,\mu\nu\rho\sigma}&=&\ft{1}{2}(g_{\mu\rho}g_{\nu\sigma}-g_{\mu\sigma}g_{\nu\rho})\,,\qquad
P_{4\partial,\mu\nu\rho\sigma}=2R_{\mu\nu\rho\sigma}\,.\label{P tensor}
\eea
(See \cite{Bueno:2016ypa} for a review of higher-derivative gravity.) For convenience, we introduce a new coordinate $x = \cos\theta \in [0,1]$ to replace the latitude angle $\theta$. The perturbative metric ansatz of the original solution \eqref{5D Kerr}, which preserves the axisymmetry, takes the form
\bea
ds^2&=&-\Big(1-\frac{2\mu}{\rho^2}-H_1\Big)dt^2
-(1+H_2)x^2\frac{4\mu b}{\rho^2}dtd\psi-(1+H_3)(1-x^2)\frac{4\mu a}{\rho^2}dtd\varphi\cr
&&+(1+H_4)x^2(1-x^2)\frac{4\mu ab}{\rho^2}d\varphi d\psi+(1+H_5)\frac{(r^2+b^2)^2(r^2+a^2x^2)-b^2r^2x^2\Delta_r}{r^2\rho^2}d\psi^2\cr
&&+(1+H_6)(1-x^2)\frac{(r^2+a^2)^2\big[r^2+b^2(1-x^2)\big]-a^2r^2(1-x^2)\Delta_r}{r^2\rho^2}d\varphi^2\cr
&&+(1+H_7)\rho^2\Big(\frac{dr^2}{\Delta_r}+d\theta^2\Big)\,.\label{5D Kerr perturbed}
\eea
Note that, compared to the $D=4$ Kerr metric, the extra dimension in the five-dimensional case corresponds to an additional Killing direction. As a result, the five-dimensional MP solution remains a cohomogeneity-2 metric. Thus, compared with the $D=4$ case in \cite{Cano:2019ore}, although the number of perturbative functions increases from four to seven, these functions still depend only on $r$ and $x$, i.e., $H_i = H_i(r,x)$. 
The fact that the metric depends only on the two-dimensional space $(r,x)$ allows us to make a gauge choice such that last term in \eqref{5D Kerr perturbed} is conformally flat.

Now, with two independent rotation parameters $a$ and $b$, we define dimensionless rotation parameters $\chi_{a,b}$
\bea
\chi_a=\frac{a}{\mu}\,,\qquad \chi_b=\frac{b}{\mu}\,.
\eea
Furthermore, we treat the two dimensionless rotation parameters as being of the same order in the rotation expansion, namely $\chi_{a,b} \sim \chi$. Consequently, the perturbative functions $H_i(r,x)$ are expanded in the power series in $\chi$:
\bea
H_i(r,x)=\sum_{n=0}^{\infty}H_i^{(n)}(r,x)\chi^n\,.\label{Hi expand 1}
\eea
As in \cite{Cano:2019ore}, we observe that $H_i^{(n)}(r,x)$ can always be expressed as a polynomial in $x$ and $1/r$
\be
H_i^{(n)}(r,x)=\sum_{p=0}^n\sum_{k=0}^{k_{\rm max}}H_i^{(n,p,k)}\frac{x^p}{r^k}\,.
\label{Hi expand 2}
\ee
In other words, for a given $n$, the polynomial in $1/r$ is of finite order with a certain maximum $k_{\rm max}$. The quantities $H_i^{(n,p,k)}$ are constant coefficients, independent of $r$ and $x$, which allows the reduction of the system of two-variable partial differential equations to a system of algebraic equations of these coefficients. A similar approach was applied to Kerr-Newman black holes \cite{Ma:2024ulp}, where the two expansion parameters corresponded to dimensionless rotation and electric charge.

In principle, $H_i$ can be expanded to all orders in $\chi$ so that \eqref{5D Kerr perturbed} solves the linearized higher-derivative equation in \eqref{quadratic}. In practice, however, we must truncate the expansion at a finite order $n$ in \eqref{Hi expand 1}, since the analytic iterative behavior is hard to determine. In this paper, we carry out the calculation up to $n=10$, and all resulting $H_i$'s are compiled in a Mathematica notebook.  Here we shall present only the first few low-order examples, up to the order $\mathcal{O}(\chi^3)$, but with the symbol omitted for simplicity. We have
\bea
H_1&=&\frac{16}{3 r^2}-\frac{16 \mu }{3 r^4}-\frac{32 \mu ^2}{3 r^6}-\frac{4}{3} (\chi _a^2-\chi _b^2) x^2 \Big(\frac{37 \mu ^2}{15
   r^4}-\frac{8 \mu ^3}{r^6}-\frac{538 \mu ^4}{15 r^8}\Big)\cr
   &&-\frac{2}{45} \Big(\frac{141 \mu  (\chi _a^2+\chi _b^2)}{r^2}-\frac{2
   \mu ^2 (59 \chi _a^2+22 \chi _b^2)}{r^4}-\frac{2 \mu ^3 (29 \chi _a^2+149 \chi _b^2)}{r^6}-\frac{1076 \mu ^4 \chi
   _b^2}{r^8}\Big)\,,\cr
H_2&=&-\frac{8}{3 r^2}-\frac{16 \mu }{3 r^4}+\frac{2}{45} (\chi _a^2-\chi _b^2) x^2 \Big(\frac{21 \mu }{r^2}+\frac{98 \mu
   ^2}{r^4}+\frac{418 \mu ^3}{r^6}\Big)\cr
   &&+\frac{2}{45} \Big(\frac{3 \mu  (17 \chi _a^2+30 \chi _b^2)}{r^2}-\frac{\mu ^2 (79 \chi
   _a^2-127 \chi _b^2)}{r^4}+\frac{418 \mu ^3 \chi _b^2}{r^6}\Big)\,,\cr
H_3&=&-\frac{8}{3 r^2}-\frac{16 \mu }{3 r^4}+\frac{2}{45} (\chi _a^2-\chi _b^2) x^2 \Big(\frac{21 \mu }{r^2}+\frac{98 \mu
   ^2}{r^4}+\frac{418 \mu ^3}{r^6}\Big)\cr
   &&+\frac{2}{45} \Big(\frac{3 \mu  (23 \chi _a^2+24 \chi _b^2)}{r^2}+\frac{\mu ^2 (29 \chi
   _a^2+19 \chi _b^2)}{r^4}+\frac{418 \mu ^3 \chi _b^2}{r^6}\Big)\,,\cr
H_4&=&-\frac{22}{3 \mu }-\frac{12}{r^2}-\frac{16 \mu }{3 r^4}+\frac{1}{45} (\chi _a^2-\chi _b^2) x^2 \Big(\frac{183 \mu }{r^2}+\frac{967
   \mu ^2}{r^4}+\frac{836 \mu ^3}{r^6}\Big)\cr
   &&+\Big(\frac{71}{20} (\chi _a^2+\chi _b^2)+\frac{\mu  (29 \chi _a^2+90 \chi
   _b^2)}{15 r^2}-\frac{\mu ^2 (158 \chi _a^2-809 \chi _b^2)}{45 r^4}+\frac{836 \mu ^3 \chi _b^2}{45 r^6}\Big)\,,\cr
H_5&=&\frac{8}{3 r^2}+\frac{4 \mu }{3 r^4}-\frac{2}{45} \Big(\frac{3 \mu  (29 \chi _a^2+18 \chi _b^2)}{r^2}+\frac{2 \mu ^2 (11 \chi
   _a^2+13 \chi _b^2)}{r^4}-\frac{22 \mu ^3 \chi _b^2}{r^6}\Big)\cr
   &&-\frac{4}{45} x^2 \Big(\frac{\mu ^2 (101 \chi _a^2+64 \chi
   _b^2)}{r^4}+\frac{2 \mu ^3 (92 \chi _a^2+73 \chi _b^2)}{r^6}+\frac{150 \mu ^4 \chi _b^2}{r^8}\Big)\,,\cr
H_6&=&\frac{8}{3 r^2}+\frac{4 \mu }{3 r^4}+\frac{4}{45} x^2 \Big(\frac{\mu ^2 (64 \chi _a^2+101 \chi _b^2)}{r^4}+\frac{2 \mu ^3 (73
   \chi _a^2+92 \chi _b^2)}{r^6}+\frac{150 \mu ^4 \chi _a^2}{r^8}\Big)\cr
   &&-\frac{2}{45} \Big(\frac{3 \mu  (18 \chi _a^2+29 \chi
   _b^2)}{r^2}+\frac{2 \mu ^3 (135 \chi _a^2+184 \chi _b^2)}{r^6}+\frac{14 \mu ^2 (11 \chi _a^2+16 \chi
   _b^2)}{r^4}+\frac{300 \mu ^4 \chi _a^2}{r^8}\Big)\,,\cr
H_7&=&\frac{8}{3 r^2}+\frac{4 \mu }{3 r^4}+\frac{2}{45} (\chi _a^2-\chi _b^2) x^2 \Big(\frac{33 \mu }{r^2}-\frac{4 \mu ^2}{r^4}+\frac{22
   \mu ^3}{r^6}\Big)\cr
   &&-\frac{2}{45} \Big(\frac{3 \mu  (29 \chi _a^2+18 \chi _b^2)}{r^2}+\frac{2 \mu ^2 (11 \chi _a^2+13 \chi
   _b^2)}{r^4}-\frac{22 \mu ^3 \chi _b^2}{r^6}\Big)\,.
\eea

\section{Black hole thermodynamic}

While the computation of thermodynamic quantities of the $D=5$ black hole is straightforward, evaluating their higher-derivative corrections based on the solution \eqref{5D Kerr perturbed} is a nontrivial task. In this section, we present a detailed, systematic procedure for computing all thermodynamic quantities.

\subsection{Asymptotic infinity and Komar charges}

The $D=5$ MP solution contains three independent parameters $(\mu, a, b)$. In our ansatz \eqref{5D Kerr perturbed} for the higher-derivative correction, we assume that these parameters remain fixed, continuing to characterize the conserved mass and two angular momenta, which can be evaluated using the Komar integration at the asymptotic infinity. It Note that all the perturbative functions $H_i$'s decay faster than the original Ricci-flat metric at the asymptotic infinity, whose structure hence remains unchanged.

The perturbative solution \eqref{5D Kerr perturbed} admits three Killing vectors, $\{\ell_t, \ell_\varphi, \ell_\psi\} = \{\partial_t, \partial_\varphi, \partial_\psi\}$, corresponding respectively to the conserved charges $\{M, J_a, J_b\}$.  Both of the two azimuthal angles in the metric \eqref{5D Kerr perturbed} are verified to have the standard $2\pi$ period. Therefore, the integration limits for the Komar integrals are given by
\bea
\int d\Omega_3=\int_{0}^{1} x dx\int_0^{2\pi}d\varphi\int_0^{2\pi}d\psi = 2\pi^2\,.
\eea
Our computations indicate that the two angular momenta
\bea
J_a=\frac{1}{16\pi}\int_{r\rightarrow\infty} {*d}\ell_\varphi=\frac{\pi}{2}\mu a\,,\qquad J_b=\frac{1}{16\pi}\int_{r\rightarrow \infty} {*d}\ell_\psi=\frac{\pi}{2}\mu b\,,
\eea
are unchanged by the higher-derivative corrections. The mass,
\bea
M=-\frac{3}{32\pi}\int_{r\rightarrow \infty} *d\ell_t =M_0+\alpha\Delta M\,,
\eea
on the other hand, receives the following correction
\bea
\Delta M&=&2 \pi -\frac{47 \pi}{20} \mu  (\chi _a^2+\chi _b^2)+\frac{\pi  \mu ^2}{280}  (401 \chi _a^4+881 \chi _a^2 \chi _b^2+401
   \chi _b^4)-\frac{\pi  \mu ^3 (\chi _a^2+\chi _b^2) }{1680}(2753 \chi _a^4\cr
&&-1412 \chi _a^2 \chi _b^2+2753 \chi_b^4)+\frac{\pi\mu ^4   }{12320}(18927 \chi _a^8+25095 \chi _a^6 \chi _b^2-12202 \chi _a^4 \chi _b^4+25095 \chi _a^2 \chi_b^6\cr
&&+18927 \chi _b^8)-\frac{\pi\mu ^5   (\chi _a^2+\chi _b^2) }{320320}(
   508397 \chi_a^8-444516 \chi _a^6 \chi_b^2+593094 \chi _a^4 \chi _b^4\cr
&&-444516 \chi _a^2 \chi _b^6+508397 \chi _b^8)\,.
\eea

\subsection{Thermodynamic quantities from the near-horizon geometry}

The calculation of the three Komar charges does not depend on the black hole horizon. In contrast, other thermodynamic quantities are directly associated with the horizon. From the metric \eqref{5D Kerr perturbed}, at this stage it is not immediately clear whether the horizon exists and remains unaffected by the higher-derivative corrections. To analyse the near-horizon geometry, we define $a_{i=1\sim8}$ and rewrite \eqref{5D Kerr perturbed} in the follow vielbein basis
\be
ds^2=a_1(d\psi+a_2d\varphi+a_3dt)^2+a_4(d\varphi+a_5dt)^2+a_6dt^2+a_7dr^2+a_8\frac{dx^2}{1-x^2}
\,.\label{5D Kerr perturbed rewrite}
\ee
Here, $a_i = a_{i,0} + \alpha \Delta a_i$, are functions of $r$ and $x$. In order to identify the horizon and compute the surface gravity, we perform a Wick rotation of the metric \eqref{5D Kerr perturbed rewrite} via \( t = i\tau \). The relevant Euclidean $\mathbb{R}^2$ part of the metric on the horizon takes the form $ ds_E^2=a_7dr^2-a_6d\tau^2+\cdots$. Specifically, it is given by
\be
ds_E^2=r^2\rho^2(1+H_7)\Big[\frac{dr^2}{r^2\Delta_r}+\frac{1+\Delta F}{r^2F}r^2\Delta_rd\tau^2
\Big]+\cdots\,,
\ee
where
\bea
F&=&2 \mu  (r^2+a^2) (r^2+b^2)+\rho ^2 r^2 \Delta _r\,,\cr
\Delta F&=&\frac{8 \mu ^2 }{\rho ^4 r^2 \Delta _r}\Big[a^2 H_3 (1-x^2) (r^2+b^2)+b^2 H_2 x^2 (r^2+a^2)\Big]-\frac{F H_1}{\rho ^2 r^2 \Delta _r}-H_7\cr
&&-\frac{4 \mu ^2}{F \rho ^4 r^4 \Delta _r}\Bigg\{
4 a^2 b^2 r^2 (r^2+a^2) (r^2+b^2) x^2(1-x^2) \mu  H_4\cr
&&+b^2 (r^2+a^2)^2 x^2 H_5
   \Big[(r^2+b^2)^2 (r^2+a^2 x^2)-b^2 r^2 x^2 \Delta _r\Big]\cr
   &&+a^2 (r^2+b^2)^2 (1-x^2) H_6
   \Big[(r^2+a^2)^2 \big[r^2+b^2 (1-x^2)\big]-a^2 r^2 (1-x^2) \Delta _r\Big]
\Bigg\}\,.
\eea
At first sight, the original horizon location with $\Delta_r (r_h)=0$ ceases to be one, due to the fact that $\Delta_r$ appears in the denominator of $\Delta F$. However, upon careful evaluation by substituting $H_i$ solutions into $\Delta F$, we find $\Delta F$ is in fact finite when $\Delta_r$ approaches zero. Therefore, the metric is degenerate on the horizon where $\Delta_r=0$. In other words, with our perturbation scheme, not only the parameters $(\mu,a,b)$ are fixed, but also the horizon location $r_h(\mu,a,b)$ remains unperturbed. It is easy to show that the surface gravity $\kappa$ on the horizon is given by
\be
\kappa^2=\frac{1+\Delta F}{4r^2F}\Big[\frac{d(r^2\Delta_r)}{dr}\Big]^2\Big|_{r=r_h}\,,
\ee
To facilitate the substitution \(r = r_h\), we express \(\mu\) in terms of $r_h$ and the expansion series of \(\chi\)
\bea
\mu&=&\frac{r_h^2}{2}+\frac{1}{8} r_h^4 (\chi _a^2+\chi _b^2)+\frac{1}{32} r_h^6 (2 \chi _a^4+5 \chi _a^2 \chi _b^2+2 \chi
   _b^4)+\frac{1}{128} r_h^8 (5 \chi _a^6+21 \chi _a^4 \chi _b^2+21 \chi _a^2 \chi _b^4+5 \chi _b^6)\cr
   &&+\frac{1}{256} r_h^{10}
   (7 \chi _a^8+42 \chi _a^6 \chi _b^2+72 \chi _a^4 \chi _b^4+42 \chi _a^2 \chi _b^6+7 \chi _b^8)+\frac{3 r_h^{12} }{2048}(14 \chi
   _a^{10}+110 \chi _a^8 \chi _b^2+275 \chi _a^6 \chi _b^4\cr
   &&+275 \chi _a^4 \chi _b^6+110 \chi _a^2 \chi _b^8+14 \chi _b^{10})\,.
\eea
In this manner, we successfully determine the corrected surface gravity,
\(\kappa^2 = \kappa_0^2 + \alpha \Delta(\kappa^2)\),
\bea
\Delta (\kappa^2)&=&-\frac{34}{3 r_h^4}+\frac{143 (\chi _a^2+\chi _b^2)}{10 r_h^2}-\frac{1}{840} (3777 \chi _a^4+5962 \chi _a^2 \chi _b^2+3777 \chi
   _b^4)+\frac{r_h^2 }{10080}(14206 \chi _a^6\cr
   &&+5835 \chi _a^4 \chi _b^2+5835 \chi _a^2 \chi _b^4+14206 \chi _b^6)-\frac{r_h^4
   }{147840}(63363 \chi _a^8+39586 \chi _a^6 \chi _b^2-249398 \chi _a^4 \chi _b^4\cr
   &&+39586 \chi _a^2 \chi _b^6+63363 \chi
   _b^8)+\frac{r_h^6 }{11531520}(2195835 \chi _a^{10}-1343891 \chi _a^8 \chi _b^2+18939144 \chi _a^6 \chi _b^4\cr
   &&+18939144 \chi _a^4 \chi
   _b^6-1343891 \chi _a^2 \chi _b^8+2195835 \chi _b^{10})\,.
\eea
We then revert \(r_h\) back to \(\mu\) and find that the temperature \(T = \kappa / (2\pi)\) is given by
\bea
T&=&T_0+\alpha\Delta T\,,\cr
\Delta T&=&-\frac{17}{12 \sqrt{2} \pi  \mu ^{3/2}}\bigg\{
1-\frac{433}{340} \mu  (\chi _a^2+\chi _b^2)+\frac{\mu ^2 }{19040}(15005 \chi _a^4+33934 \chi _a^2 \chi _b^2+15005 \chi
   _b^4)\cr
   &&-\frac{\mu ^3 (\chi _a^2+\chi _b^2) }{228480}(219493 \chi _a^4-200794 \chi _a^2 \chi _b^2+219493 \chi
   _b^4)+\frac{\mu ^4 }{40212480}(35245889 \chi _a^8\cr
   &&+52238684 \chi _a^6 \chi _b^2-22507578 \chi _a^4 \chi _b^4+52238684 \chi _a^2 \chi
   _b^6+35245889 \chi _b^8)\cr
   &&-\frac{\mu ^5 (\chi _a^2+\chi _b^2) }{2091048960}(1923648557 \chi _a^8-2529489028 \chi _a^6 \chi
   _b^2+2776653870 \chi _a^4 \chi _b^4\cr
   &&-2529489028 \chi _a^2 \chi _b^6 +1923648557 \chi _b^8)
\bigg\}\,.
\eea
Having determined that the horizon exists, unperturbed by the higher-derivative correction, then the two angular velocities can be expressed as
\bea
\Omega_a&=&a_5|_{r=r_h}^{r\rightarrow\infty}\,,\qquad \Omega_b=(a_3-a_2a_5)|_{r=r_h}^{r\rightarrow\infty}\,.
\eea
The higher-derivative corrections to them are given by
\bea
\Omega_{a,b}&=&\Omega_{a,b,0}+\alpha\Delta\Omega_{a,b}\,,\cr
\Delta\Omega_{a}&=&-\frac{13 \chi _a}{6 \mu }+\frac{1}{60} \chi _a (123 \chi _a^2+163 \chi _b^2)-\frac{1}{840} \mu  \chi _a (1779 \chi _a^4+1424
   \chi _a^2 \chi _b^2-1126 \chi _b^4)\cr
   &&+\frac{\mu ^2 \chi _a }{5040}(10487 \chi _a^6+4191 \chi _a^4 \chi _b^2-1032 \chi _a^2 \chi
   _b^4+11624 \chi _b^6)-\frac{\mu ^3 \chi _a }{110880}(232899 \chi _a^8\cr
   &&+49724 \chi _a^6 \chi _b^2-232884 \chi _a^4 \chi _b^4+203130
   \chi _a^2 \chi _b^6-48547 \chi _b^8)\,,\cr
   \Delta\Omega_{b}&=&\Delta\Omega_{a}|_{a\leftrightarrow b}\,.
\eea

Of all physical quantities associated with the black hole horizon, the entropy in our perturbed solution is the most challenging to compute. In higher-derivative gravity theories, the entropy is no longer simply proportional to the horizon area, it is instead determined by the Wald entropy formula \cite{Wald:1993nt}
\bea
S=-\frac{1}{8}\int_{S^3} d\Omega_3\sqrt{h}P_{\mu\nu\rho\sigma}\epsilon^{\mu\nu}\epsilon^{\rho\sigma}\big|_{r=r_h}
\eea
Here, \(h\) denotes the determinant of the three-dimensional metric obtained by fixing \(t\) and \(r\), and \(\epsilon^{\mu\nu}\) is the binormal vector of the black hole horizon.

In principle, the entropy could be computed directly from the above expression. However, the complexity of the perturbed solution \eqref{5D Kerr perturbed} renders an explicit evaluation practically unmanageable. We therefore need to adopt a more refined strategy. By examining the tensor \(P_{\mu\nu\rho\sigma}\) in \eqref{P tensor}, we observe that \(P_{4\partial,\mu\nu\rho\sigma}\) itself is already an \(\alpha\)-order correction. Consequently, when evaluating its contribution to the black hole entropy, it suffices to substitute only the leading-order solution \eqref{5D Kerr}
\bea
\Delta S_{1}=\frac{2 \pi ^2 }{r_h^3}\big[3 r_h^4+(a^2+b^2) r_h^2-a^2 b^2\big]\,.
\eea
In contrast, the contribution of \(P_{E,\mu\nu\rho\sigma}\) to the entropy is simply one quarter of the horizon area, which can be read directly from \eqref{5D Kerr perturbed rewrite}
\bea
A=\int d\Omega_3\sqrt{a_1a_4\frac{a_8}{1-x^2}}\,.
\eea
By expanding, we can obtain the contribution of the area part to the higher-order corrections of the black hole entropy
\bea
\frac{A}{4}&=&S_0+\alpha\Delta S_2\,,\cr
\Delta S_2&=&\frac{5\pi ^2   r_h}{2} \Big[
1-\frac{16}{25} r_h^2 (\chi _a^2+\chi _b^2)-\frac{r_h^4 }{2800}(59 \chi _a^4-285 \chi _a^2 \chi _b^2+59 \chi
   _b^4)-\frac{r_h^6 (\chi _a^2+\chi _b^2) }{8400}(1147 \chi _a^4\cr
   &&-1195 \chi _a^2 \chi _b^2+1147 \chi
   _b^4)-\frac{r_h^8 }{1478400}(81314 \chi _a^8+57935 \chi _a^6 \chi _b^2+42654 \chi _a^4 \chi _b^4+57935 \chi _a^2 \chi _b^6\cr
   &&+81314
   \chi _b^8)-\frac{r_h^{10} (\chi _a^2+\chi _b^2) }{19219200}(1102396 \chi _a^8+78581 \chi _a^6 \chi _b^2+849965 \chi
   _a^4 \chi _b^4+78581 \chi _a^2 \chi _b^6\cr
   &&+1102396 \chi _b^8)
\Big]\,.
\eea
The total higher-derivative correction to the entropy is given by the sum of the two contributions, \(\Delta S_1\) and \(\Delta S_2\). Rewriting all instances of \(r_h\) in \(\Delta S_{1,2}\) in terms of \(\mu\) yields the final result
\bea
\Delta S&=&\frac{17 \pi    \sqrt{\mu }}{\sqrt{2}}\Big[
1-\frac{173}{340} \mu  (\chi _a^2+\chi _b^2)+\frac{3 \mu ^2 }{19040}(1623 \chi _a^4+3154 \chi _a^2 \chi _b^2+1623 \chi
   _b^4)-\frac{\mu ^3  }{228480}(\chi _a^2\cr
   &&+\chi _b^2)(73657 \chi _a^4-31738 \chi _a^2 \chi _b^2+73657 \chi
   _b^4)+\frac{\mu ^4 }{40212480}(11702545 \chi _a^8+14760364 \chi _a^6 \chi _b^2\cr
   &&-12042618 \chi _a^4 \chi _b^4+14760364 \chi _a^2 \chi
   _b^6+11702545 \chi _b^8)-\frac{\mu ^5 (\chi _a^2+\chi _b^2) }{697016320}(214206659 \chi _a^8\cr
   &&-206128620 \chi _a^6 \chi
   _b^2+331260242 \chi _a^4 \chi _b^4-206128620 \chi _a^2 \chi _b^6+214206659 \chi _b^8)
\Big]\,.
\eea
We have computed all the thermodynamic quantities, and upon verification, the higher-derivative corrected quantities satisfy the first law of black hole thermodynamics
\be
dM=T dS+\Omega_{a} dJ_{a}+\Omega_{b} dJ_{b}+\mathcal{O}(\alpha^2)\,.\label{first law}
\ee
In appendix, we demonstrate that these thermodynamic quantities are consistent with those obtained from the RS method.

\subsection{Parameter redefinition}

So far, we have fully computed all the black hole thermodynamic quantities and verified that they satisfy the first law \eqref{first law}, providing a nontrivial validation of the perturbed solution \eqref{5D Kerr perturbed}. As mentioned, in our perturbative scheme, the original MP parameters $(\mu,a,b)$ and the horizon remain fixed under the higher-derivative perturbation. However, these are not direct physical quantities. It is more instructive to consider corrections for a set of fixed three basic variables. Here we consider perturbations with fixed conserved quantities, namely the mass and two angular momenta.  To do so, we redefine the three original parameters \(\{\mu, \chi_a, \chi_b\}\)
\be
\mu\rightarrow\mu'=\mu+\alpha\delta\mu\,,\qquad  \chi_a\rightarrow\chi_a'=\chi_a+\alpha\delta\chi_a
\,,\qquad  \chi_b\rightarrow\chi_b'=\chi_b+\alpha\delta\chi_b\,,\label{parameter redefinition}
\ee
so that the mass $M$ and angular momenta $J_{a,b}$ are fixed
\bea
M=M_0+\mathcal{O}(\alpha^2)\,,\qquad J_a=J_{a,0}+\mathcal{O}(\alpha^2)\,,\qquad J_b=J_{b,0}+\mathcal{O}(\alpha^2)\,.\label{fix charges}
\eea
We have
\bea
\delta\mu&=&-\frac{8}{3}+\frac{47\mu}{15}   (\chi _a^2+\chi _b^2)-\frac{\mu ^2}{210}  (401 \chi _a^4+881 \chi _a^2 \chi _b^2+401 \chi
   _b^4)+\frac{\mu ^3 (\chi _a^2+\chi _b^2) }{1260}(2753 \chi _a^4\cr
   &&-1412 \chi _a^2 \chi _b^2+2753 \chi
   _b^4)-\frac{\mu ^4 }{9240}(18927 \chi _a^8+25095 \chi _a^6 \chi _b^2-12202 \chi _a^4 \chi _b^4+25095 \chi _a^2 \chi _b^6\cr
   &&+18927
   \chi _b^8)+\frac{\mu ^5 (\chi _a^2+\chi _b^2) }{240240}(508397 \chi _a^8-444516 \chi _a^6 \chi _b^2+593094 \chi _a^4 \chi
   _b^4-444516 \chi _a^2 \chi _b^6\cr
   &&+508397 \chi _b^8)\,,\cr
\delta\chi_a&=&\frac{16 \chi _a}{3 \mu }-\frac{94\chi _a}{15}  (\chi _a^2+\chi _b^2)+\frac{\mu  \chi _a}{105}  (401 \chi _a^4+881 \chi _a^2
   \chi _b^2+401 \chi _b^4)-\frac{\mu ^2 \chi _a}{630}  (\chi _a^2+\chi _b^2) (2753 \chi _a^4\cr
   &&-1412 \chi _a^2 \chi
   _b^2+2753 \chi _b^4)+\frac{\mu ^3 \chi _a }{4620}(18927 \chi _a^8+25095 \chi _a^6 \chi _b^2-12202 \chi _a^4 \chi _b^4+25095 \chi _a^2
   \chi _b^6\cr
   &&+18927 \chi _b^8)\,,\cr
\delta\chi_b&=&\delta\chi_a|_{a\leftrightarrow b}\,.
\eea
Consequently, the higher-order corrections to the temperature \(T\), entropy $S$ and the angular velocities \(\Omega_{a,b}\) become
\bea
T=T_0+\alpha\Delta T\,,\qquad S=S_0+\alpha\Delta S\,,\qquad \Omega_{a}=\Omega_{a,0}+\alpha\Delta \Omega_{a}\,,\qquad \Omega_{b}=\Omega_{b,0}+\alpha\Delta \Omega_{b}
\eea
are given by
\bea
\Delta T&=&-\frac{3  }{4 \sqrt{2} \pi  \mu ^{3/2}}\Big[
1+\frac{7\mu}{36}   (\chi _a^2+\chi _b^2)-\frac{\mu ^2}{288}  (177 \chi _a^4-970 \chi _a^2 \chi _b^2+177 \chi
   _b^4)\cr
&&-\frac{\mu ^3 (\chi _a^2+\chi _b^2) }{1152}(305 \chi _a^4 -3234 \chi _a^2 \chi _b^2+305 \chi _b^4)-\frac{\mu
   ^4 }{18432}(2653 \chi _a^8-32500 \chi _a^6 \chi _b^2\cr
&&-103378 \chi _a^4 \chi _b^4-32500 \chi _a^2 \chi _b^6 +2653 \chi
   _b^8)-\frac{\mu ^5 (\chi _a^2+\chi _b^2) }{73728}(5823 \chi _a^8-83660 \chi _a^6 \chi _b^2\cr
&&-466022 \chi _a^4 \chi
   _b^4-83660 \chi _a^2 \chi _b^6+5823 \chi _b^8)
\Big]\,,\nn\\
\Delta S&=&\frac{9 \pi ^2   \sqrt{\mu }}{\sqrt{2}}\Big[
1-\frac{5\mu}{36}   (\chi _a^2+\chi _b^2)+\frac{7\mu ^2}{288}  (\chi _a^4-18 \chi _a^2
\chi_b^2+\chi _b^4)+\frac{\mu ^3}{128}
    (3 \chi _a^6-43 \chi _a^4 \chi _b^2\cr
&&-43 \chi _a^2 \chi _b^4 +3 \chi _b^6)+\frac{11 \mu ^4 }{18432}(25 \chi _a^8-340 \chi _a^6
   \chi _b^2-1162 \chi _a^4 \chi _b^4-340 \chi _a^2 \chi _b^6+25 \chi _b^8)\cr
&&+\frac{91 \mu ^5 }{73728}(7 \chi _a^{10}-85 \chi _a^8 \chi_b^2 -690 \chi _a^6 \chi _b^4-690
\chi_a^4 \chi _b^6-85 \chi _a^2 \chi _b^8+7 \chi _b^{10})
\Big]\,,\cr
\Delta \Omega_{a}&=&\frac{\chi _a}{2 \mu }-\frac{\chi _a}{12}  (13 \chi _a^2-35 \chi _b^2)-\frac{\mu  \chi _a}{24}  (5 \chi _a^4-30 \chi _a^2 \chi
   _b^2-48 \chi _b^4)-\frac{\mu ^2 \chi _a}{48}  (5 \chi _a^6-41 \chi _a^4 \chi _b^2\cr
&&-154 \chi _a^2 \chi _b^4-58 \chi_b^6)-\frac{\mu ^3 \chi _a}{96}  (5 \chi _a^8-44 \chi _a^6
\chi _b^2-378 \chi _a^4 \chi _b^4-388 \chi _a^2 \chi _b^6-65 \chi_b^8)\,,\cr
\Delta \Omega_{b}&=&\Delta \Omega_{a}|_{a\leftrightarrow b}\,.
\eea
The correction to the Gibbs free energy with these variables,
\be
G=G_0+\alpha\Delta G\,,
\ee
can then be computed, given  by
\bea
\Delta G&=&-\frac{3 \pi   }{2}\Big[
1-\frac{7\mu}{18}   (\chi _a^2+\chi _b^2)+\frac{\mu ^2}{36}  (\chi _a^4-29 \chi _a^2 \chi _b^2+\chi _b^4)
+\frac{\mu ^3(\chi _a^2+\chi _b^2)}{72}   (\chi _a^4-35 \chi _a^2 \chi
   _b^2+\chi _b^4)\cr
   &&+\frac{\mu ^5(\chi _a^2+\chi
   _b^2)}{288}   (\chi _a^8-42 \chi _a^6 \chi _b^2-188 \chi _a^4 \chi _b^4-42 \chi _a^2 \chi _b^6+\chi _b^8)+\frac{1}{144} \mu ^4
   (\chi _a^8-38 \chi _a^6 \chi _b^2\cr
   &&-108 \chi _a^4 \chi _b^4-38 \chi _a^2 \chi _b^6+\chi _b^8)
\Big]\,.
\eea

\section{Conclusion}

In this paper, we considered quadratic curvature perturbation to the five-dimensional MP black hole at the linear level in the coupling constant $\alpha$. The metric is cohomogeneity two, depending on the radial and latitude coordinates. The solution is solved order by order in terms of two dimensionless angular momentum parameters. In principle, the solution can be solved up to an arbitrary order, and we presented the results up to the tenth order. In this perturbation scheme, the original MP parameters $(\mu,a,b)$ are chosen to be fixed and intriguingly, the horizon location also remains fixed. The solution allowed us to compute the higher-derivative corrections to the black hole thermodynamics of the five-dimensional MP solution. We found that the results were in precise agreement with those from the RS method.

There are certain limitations in this perturbative approach, even though we can achieve arbitrary higher orders in the dimensionless angular momentum expansion, making the solution potentially valid for large angular momenta. The perturbative functions are all analytic in $r$ and $x=\cos\theta$ coordinates. This analytic perturbative approach 
does not capture the nontrivial geometric structure that may emerge in extremal black holes, where the horizon can be singular and characterized by irrational exponents in near-horizon geometry expansion, such as $(r-r_0)^\Delta$ with irrational $\Delta$ \cite{Mao:2023qxq,Mao:2025nrp}. This method will also become increasingly unmanageable in higher dimensions when the metric becomes increasingly higher cohomogeneity.

Nevertheless, out solution has nontrivial applications. It allows us to derive the higher-derivative correction to thermodynamics at the NNLO \cite{Ma:2023qqj}. One can also study higher-derivative corrections to quasi-normal modes, multipole moments, etc., which
rely heavily on the explicit expressions of the solution. These are among the topics that are worth further investigation.

\section*{Acknowledgement}

L.M.~is supported in part by National Natural Science Foundation of China (NSFC) grant No.~12447138, Postdoctoral Fellowship Program of CPSF Grant No.~GZC20241211, the China Postdoctoral Science Foundation under Grant No.~2024M762338 and the National Key Research and Development Program No.~2022YFE0134300. H.L.~is supported in part by the NSFC grants No.~12375052 and No.~11935009. The work is also supported in part by the Tianjin University Self-Innovation Fund Extreme Basic Research Project Grant No.~2025XJ21-0007.

\appendix

\section{Higher-derivative corrections from RS method}

In recent years, the study of thermodynamic corrections to rotating black holes within effective field theory has attracted significant attention. Based on the Euclidean path integral approach \cite{Gibbons:1976ue}, Reall and Santos proposed the RS method \cite{Reall:2019sah}. This method allows us to substitute the leading-order exact solution directly into the higher-derivative correction terms, yielding the complete free energy. By avoiding the need to solve the complicated equations of motion for higher-order corrections, the RS method has become widely adopted among researchers.

While the RS method has been highly successful, this does not diminish the importance of obtaining explicit solutions with higher-order corrections, as explicit solutions can have applications beyond thermodynamics. Furthermore, the RS method assumes that the black holes continue to exist under higher-derivative correction, which requires justification. In this appendix, we gives the RS method result and compare it with our thermodynamic quantities.

In \cite{Wu:2024iiz}, the RS method was employed to compute the the thermodynamic corrections of MP black holes in general dimensions under the quadratic curvature correction. In \cite{Ma:2024ynp,Chen:2025ary}, the RS method was employed to compute the thermodynamic corrections of five-dimensional rotating AdS black holes. Specifically, when the temperature \(T\) and angular velocities \(\Omega_{a,b}\) are held fixed under the higher-order corrections, the Gibbs free energy acquires the following higher-order correction
\bea
\alpha\Delta G(T,\Omega_a,\Omega_b,\alpha)&=&-\frac{\pi  \alpha }{4 r_h^4 (r_h^2+a^2) (r_h^2+b^2)}\Big[
a^4 b^4-6 a^2 b^2 (a^2+b^2) r_h^2\cr
&&+(a^4-20 a^2 b^2+b^4) r_h^4+2 (a^2+b^2) r_h^6+9 r_h^8
\Big],\label{RS 1}
\eea
and then we expand it in powers of \(\chi\) up to \(\chi^{10}\)
\bea
\alpha\Delta G&=&-\frac{9 \pi  \alpha }{4}\Big[
1-\frac{7\mu}{18}   (\chi _a^2+\chi _b^2)+\frac{\mu ^2 }{36} (\chi _a^4
-29 \chi _a^2 \chi_b^2+\chi _b^4)+\frac{\mu ^3 (\chi _a^2+\chi _b^2)}{72}  (\chi _a^4-35 \chi _a^2 \chi_b^2\cr
&&+\chi _b^4)+\frac{\mu ^5 (\chi _a^2+\chi_b^2)}{288}  (\chi _a^8-42 \chi _a^6 \chi _b^2
-188 \chi _a^4 \chi _b^4-42 \chi _a^2 \chi _b^6+\chi _b^8)+\frac{\mu ^4}{144}
(\chi _a^8\cr
&&-38 \chi _a^6 \chi _b^2-108 \chi _a^4 \chi _b^4-38 \chi _a^2 \chi _b^6+\chi _b^8)
\Big]\,.\label{RS 2}
\eea
To compare the thermodynamic quantities from our perturbed solution \eqref{5D Kerr perturbed}, as obtained in sections 3.1 and 3.2, with those from the RS method \eqref{RS 2}, we perform a parameter redefinition that preserves the temperature \(T\) and angular velocities \(\Omega_{a,b}\)
\bea
T=T_0+\mathcal{O}(\alpha^2)\,,\qquad \Omega_a=\Omega_{a,0}+\mathcal{O}(\alpha^2)\,,\qquad \Omega_b=\Omega_{b,0}+\mathcal{O}(\alpha^2)\,.\label{fix potential}
\eea
The required parameter redefinition \eqref{parameter redefinition} can then be determined
\bea
\delta\mu&=&-\frac{17}{3}+\frac{43\mu}{10}   (\chi _a^2+\chi _b^2)-\frac{3\mu ^2}{140} (93 \chi _a^4+83 \chi _a^2 \chi _b^2+93 \chi
   _b^4)+\frac{\mu ^3 (\chi _a^2+\chi _b^2) }{2520}(5401 \chi _a^4\cr
&&+851 \chi _a^2 \chi _b^2 +5401 \chi _b^4)-\frac{\mu
   ^4 }{18480}(38239 \chi _a^8+35560 \chi _a^6 \chi _b^2-65984 \chi _a^4 \chi _b^4+35560 \chi _a^2 \chi _b^6\cr
&&+38239 \chi_b^8) +\frac{\mu ^5(\chi _a^2+\chi _b^2) }{480480}(1011789 \chi _a^8-678822 \chi _a^6 \chi _b^2+2127128 \chi _a^4 \chi_b^4\cr
&&-678822 \chi _a^2 \chi _b^6+1011789 \chi _b^8)\,,\cr
\delta\chi_a&=&\frac{13 \chi _a}{3 \mu }-\frac{\chi _a}{10} (41 \chi _a^2+91 \chi _b^2)+\frac{\mu  \chi _a}{140}  (593 \chi _a^4+1373 \chi
   _a^2 \chi _b^2+173 \chi _b^4)-\frac{\mu ^2 \chi _a }{2520}(10487 \chi _a^6\cr
   &&+534 \chi _a^4 \chi _b^2+9354 \chi _a^2 \chi _b^4+14897 \chi
   _b^6)+\frac{\mu ^3 \chi _a }{18480}(77633 \chi _a^8+133490 \chi _a^6 \chi _b^2\cr
   &&-34948 \chi _a^4 \chi _b^4 +41090 \chi _a^2 \chi
   _b^6+59153 \chi _b^8)  \,,\cr
   \delta\chi_b&=&\delta\chi_a|_{a\leftrightarrow b}\,.
\eea

Under these circumstances, the black hole mass \(M\), entropy \(S\), and angular momenta \(J_{a,b}\) acquire the corresponding corrections
\bea
M=M_0+\alpha\Delta M\,,\qquad S=S_0+\alpha\Delta S\,,\qquad J_{a}=J_{a,0}+\alpha\Delta J_{a}\,,\qquad J_{b}=J_{b,0}+\alpha\Delta J_{b}\,.
\eea
They are
\bea
\Delta M&=&-\frac{9 \pi}{4}+\frac{7\pi  \mu}{8}  (\chi _a^2+\chi _b^2)-\frac{\pi  \mu ^2}{16} (\chi _a^4-29 \chi _a^2 \chi _b^2+\chi
   _b^4)-\frac{\pi  \mu ^3(\chi _a^2+\chi _b^2)}{32}
   (\chi _a^4-35 \chi _a^2 \chi _b^2+\chi _b^4)\cr
   &&-\frac{\pi  \mu ^5 (\chi _a^2+\chi _b^2)}{128}  (\chi _a^8-42 \chi _a^6 \chi _b^2-188 \chi _a^4 \chi _b^4-42 \chi
   _a^2 \chi _b^6+\chi _b^8)-\frac{\pi  \mu ^4}{64} (\chi _a^8-38 \chi _a^6 \chi _b^2\cr
   &&-108 \chi _a^4 \chi _b^4-38 \chi _a^2 \chi
   _b^6+\chi _b^8)\,,\cr
\Delta S&=&\frac{\pi ^2 \mu ^{3/2}}{\sqrt{2}}\Big[
7 (\chi _a^2+\chi _b^2)-\frac{\mu}{4} (11 \chi _a^4-46 \chi _a^2 \chi _b^2+11 \chi _b^4)-\frac{\mu ^2 (\chi
   _a^2+\chi _b^2)}{32} (23 \chi _a^4-134 \chi _a^2 \chi _b^2\cr
&&+23 \chi _b^4)-\frac{\mu ^3}{128}  (43 \chi _a^8-116 \chi _a^6
   \chi _b^2-686 \chi _a^4 \chi _b^4-116 \chi _a^2 \chi _b^6+43 \chi _b^8)-\frac{5 \mu ^4 (\chi _a^2+\chi _b^2) }{2048}(71 \chi
   _a^8\cr
&&+4 \chi _a^6 \chi _b^2-1942 \chi _a^4 \chi _b^4+4 \chi _a^2 \chi _b^6+71 \chi _b^8)
\Big]\,,\cr
\Delta J_{a}&=&-\frac{7}{2} \pi  \mu  \chi _a\Big[
1-\frac{\mu}{14}  (9 \chi _a^2-\chi _b^2)-\frac{\mu ^2}{28} (\chi _a^4+25 \chi _a^2 \chi _b^2-11 \chi
   _b^4)-\frac{\mu ^3}{56}  (\chi _a^6+38 \chi _a^4 \chi _b^2+10 \chi _a^2 \chi _b^4\cr
   &&-13 \chi _b^6)-\frac{\mu ^4}{112}
   (\chi _a^8+54 \chi _a^6 \chi _b^2+84 \chi _a^4 \chi _b^4-26 \chi _a^2 \chi _b^6-15 \chi _b^8)
\Big]\,,\cr
\Delta J_{b}&=&\Delta J_{a}|_{a\leftrightarrow b}\,.
\eea
We find that the higher-order corrections to the Gibbs free energy obtained from the above thermodynamic quantities are in complete agreement with the results from the RS method \eqref{RS 2}.

\end{document}